\documentclass[%
 aip,
 amsmath,amssymb,
 reprint,%
]{revtex4-1}

\usepackage{graphicx}
\usepackage{epsfig}
\usepackage{epstopdf}
\usepackage{subfigure,color}
\usepackage{dcolumn}
\usepackage{bm}
\usepackage{lineno}
\usepackage{comment}
\usepackage{graphicx}
\usepackage{dcolumn}
\usepackage{bm}

\usepackage[utf8]{inputenc}
\usepackage[T1]{fontenc}
\usepackage{mathptmx}
\usepackage{etoolbox}

\makeatletter
\def\@email#1#2{%
 \endgroup
 \patchcmd{\titleblock@produce}
  {\frontmatter@RRAPformat}
  {\frontmatter@RRAPformat{\produce@RRAP{*#1\href{mailto:#2}{#2}}}\frontmatter@RRAPformat}
  {}{}
}%
\makeatother
\begin{document}

\preprint{AIP/123-QED}

\title{Antenna bandwidth engineering through time-varying resistance}

\affiliation{
Department of Electronics and Nanoengineering, Aalto University, P.O.~Box 15500, FI-00076 Aalto, Finland}
\email{mohamed.mostafa@aalto.fi}

\author{M.~H.~Mostafa$^*$ }%
\author{N.~Ha-Van}%
\author{P.~Jayathurathnage}%
\author{X.~Wang}%
\author{G.~Ptitcyn}%
\author{S.~A.~Tretyakov}%



\begin{abstract}
Operational bandwidth of resonant circuits is limited by the resonator's size, which is known as the Chu limit. This limit restricts miniaturization of antennas, as the antenna bandwidth is inversely proportional to its size. Here, we propose slow time modulation of resistive elements to engineer bandwidth of small antennas. The temporal modulation of resistance induces virtual impedance that is fully controlled by the modulation parameters. We show how the virtual impedance can be used to optimize the frequency response of a resonant circuit, leading to enhanced matching at multiple frequencies simultaneously. We experimentally verify the proposed technique, demonstrating enhancement of radiation of a broadband modulated signal radiated by a small antenna.
\end{abstract}

\maketitle 





Performance of small antennas is fundamentally limited by   Chu's and Fano's  limits~\cite{chu1948physical,wheeler1947fundamental,FANO1950139}. Chu's limit defines a  trade-off between the bandwidth and size of passive, linear, and time-invariant antennas. In order to overcome Chu's limit one needs to utilize either active, nonlinear~\cite{hartley1936oscillations}, or time-varying systems~\cite{jacob1954keying,wolff1957high,galejs1963electronic,galejs1963switching,hartley1968analysis}. Currently, there is growing interest towards leveraging time-varying circuit elements to enhance bandwidth of small antennas. It has been shown that matching networks with time-varying reactive elements can break Chu's limit~\cite{Mekawy2019, PhysRevApplied.15.054063,Hrabar2020}, while the use of time-varying transmission line parameters allows to overcome the Bode-Fano bound~\cite{Yakir2018, 8966563, yang2022broadband}. Similarly, temporal modulation of thin absorbers induces spectrum spreading, where the scattered fields can be maintained below a certain level, extending absorber's bandwidth~\cite{Chambers2005, caloz2021Camouflaging, Huanan2021}. Furthermore, direct antenna modulation is also a possibility to overcome bandwidth limitations~\cite{DAM1, DAM2, DAM3, DAM4, DAM5}. Most of the known approaches exploit time-varying reactive elements, which are prone to instabilities. Other approaches are based on the use of switches, allowing only pulse modulation. 

Recently, double-frequency modulation of resistive elements was introduced to enhance bandwidth of small antennas~\cite{PhysRevApplied.16.014017}. In addition, it was shown that for multifrequency input signals, slow modulations of resistive elements can be utilized to achieve perfect impedance matching at multiple frequencies simultaneously~\cite{PhysRevApplied.17.064048}. This method of bandwidth engineering~\cite{PhysRevApplied.16.014017, PhysRevApplied.17.064048} stems from the equivalence of a time-varying resistor to a static resistor in series with a virtual impedance element that can be fully reactive. By designing the modulation, the virtual reactive impedance can be controlled at every frequency independently, providing an intriguing opportunity to fully engineer the frequency response. Slowly varying resistive elements have multiple advantages. Firstly, slow modulation is practically more easy to realize than the double-frequency modulation. Secondly, when the resistance profile is kept always positive, the system is unconditionally stable, in contrast to modulated reactive elements. Finally, as the resistance profile is always positive, there is no power exchange between the main circuit and the modulation circuit.

In this paper, we build on previous works~\cite{PhysRevApplied.16.014017, PhysRevApplied.17.064048} and investigate a possibility of using slowly varying resistive elements to enhance the matching of a narrowband resonant transmitting antenna for all harmonics of multifrequency input signals. We derive expressions for the virtual impedances induced due to the modulation, and apply this technique to enhance radiation from a small antenna at multiple frequencies simultaneously. Finally, we verify the validity of this technique experimentally, where the power radiated by a planar inverted-F antenna (PIFA) is enhanced by leveraging a dynamic resistive element.

\begin{figure*}
\centerline{\includegraphics[width=1\linewidth]{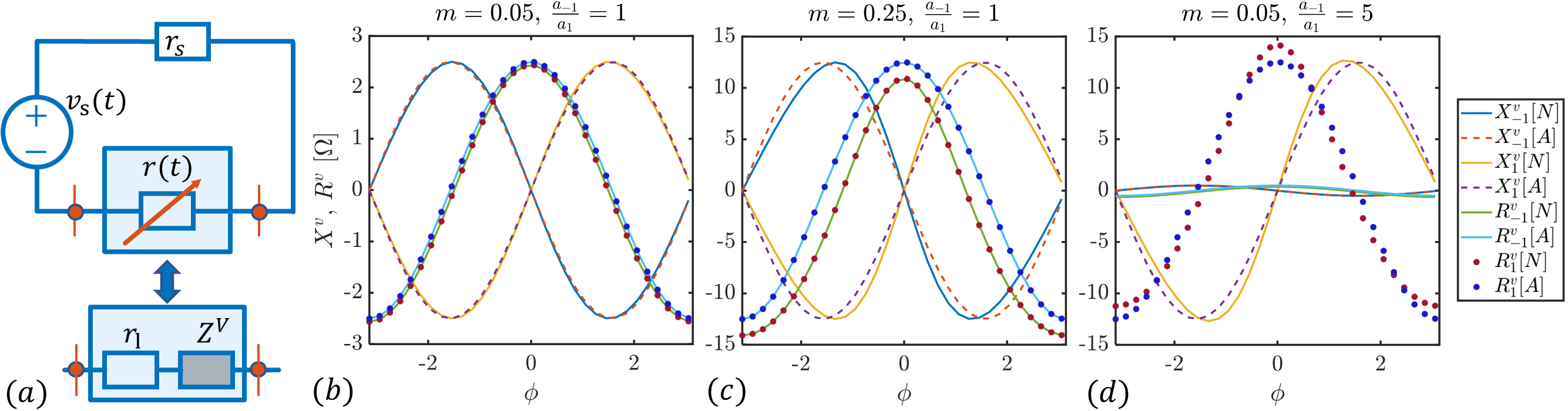}}
\caption{(a) Circuit schematic. (b)-(d) Real and imaginary parts of the virtual impedance $Z^{\rm{v}}_{-1,1}=R^{\rm{v}}_{-1,1}+jX^{\rm{v}}_{-1,1}$ as function of the modulation phase $\phi$ for $\theta_{-1,1}=0$, where $-1, 1$ refers to $\omega_{-1}, \omega_{1}$. Analytical results are marked by [A] and numerical results are marked by [N].}\label{fig1}
\end{figure*}

Let us consider a voltage source $v_{\rm s}$ connected to a time-varying resistor as shown in Fig.~\ref{fig1}(a). The applied voltage contains two frequency compoenents, and the resistance is modulated at the difference frequency: \begin{equation}
\begin{split}
v_\text{s}(t) & =a_{-1}\cos{(\omega_{-1}t+\theta_{-1})}+ a_{1}\cos{(\omega_1t+\theta_{1})},\cr
r_{\rm t}(t) & =r_0 + r_0 m \cos{(\Delta \omega t+\phi)=r_{\rm s}+r(t)}.
\end{split}
\label{eq:volsou1}
\end{equation}
Here, $a_n$, $\omega_{n}$, $\theta_{n}$, $r_0$, $m$ are the voltage amplitude, angular frequency, phase, total static resistance $r_0=r_{\rm s}+r_{\rm l}$, and the  modulation amplitude, respectively. The total static resistance is composed by the source resistance $r_{\rm s}$ and the static component in the varying resistance $r_{\rm l}$. $\Delta \omega=\omega_1 - \omega_{-1}$ is the modulation angular frequency, and $\phi$ denotes the modulation phase. 
The modulation amplitude is assumed to be much smaller than unity, and we use the approximation  $(1+x)^{-1}\approx1-x$ in order to calculate the current. 
Two current components have angular frequencies different from $\omega_{-1}$ and $\omega_{1}$, and we assume that they are filtered out. In this case, the electric current is simplified and can be written as 
\begin{equation}
\begin{split}
&i(t)\approx\cr
&{1\over r_0}\bigg[a_{-1}\cos(\omega_{-1}t+\theta_{-1})-{m\over2}a_{1}\cos(\omega_{-1}t+\theta_{1}-\phi)\cr
& + a_{1}\cos(\omega_{1}t+\theta_{1})-{m\over2}a_{-1}\cos(\omega_{1}t+\theta_{-1}+\phi)\bigg].
\end{split}
\label{eq:curmohem}
\end{equation}
Let us use the phasor notation and write the electric current, for example, at only $\omega_{1}$:
\begin{equation}
I_{\omega_{1}}={a_1\over r_0}e^{j\theta_1}-{m\over2r_0}a_{-1}e^{j(\theta_{-1}+\phi)}.   
\end{equation}
On the other hand, the  complex voltage amplitude at  $\omega_1$ is simply $V_{\omega_1}=a_1e^{j\theta_1}$. In consequence, the corresponding ratio of the amplitudes of voltage and current reads
\begin{equation}
{V_{\omega_{1}}\over I_{\omega_{1}}}=r_0\Big(1-{m\over2}{a_{-1}\over a_{1}}e^{j(\theta_{-1}+\phi-\theta_1)}\Big)^{-1}.\label{VI_omega1}
\end{equation}

Assuming that $m{a_{-1}\over a_1}\ll 1$,
we find the equivalent impedance at this frequency as 
\begin{equation}
Z^{\rm{eq}}|_{\omega=\omega_1}\approx r_0+r_0{m\over2}{a_{-1}\over a_{1}}e^{j(\theta_{-1}+\phi-\theta_1)}.
\label{eq:intint}
\end{equation}
We see that under this assumption the time-varying resistor is equivalent to $r_0$ connected in series with a virtual impedance $Z^{\rm{v}}$, given by the second term in \eqref{eq:intint}. This equivalent model of the circuit is shown in Fig.~\ref{fig1}(a). Depending on the modulation phase $\phi$, this virtual impedance can be purely real, purely imaginary, or complex-valued.
The same procedure can be done for the other angular frequency, resulting in
\begin{equation}
Z^{\rm{eq}}|_{\omega=\omega_{-1}}\approx r_0+r_0{m\over2}{a_{1}\over a_{-1}}e^{j(\theta_{1}-\phi-\theta_{-1})}.
\label{eq:intint1}
\end{equation}
By properly selecting the modulation amplitude and phase, we can engineer these virtual impedances. Considering $Z^{\rm{v}}|_{\omega=\omega_1}$, if 1)~$\theta_{-1}+\phi-\theta_1=h\pi$, we have an extra resistance such that for even values of $h$ we add more loss to the circuit and for odd values of $h$, we add gain (negative resistance). On the other hand, if 2)~$\theta_{-1}+\phi-\theta_1=(2h+1)\pi/2$, the virtual impedance turns out to be purely imaginary, corresponding to a reactance. If $h$ is even, this reactance is inductive, and if $h$ is odd, the reactance is capacitive. 





We confirm the analytical results numerically using the impedance-matrix method~\cite{Wang}. Let us assume that the angular frequencies are $\omega_{-1}=0.9\cdot 10^{9}\cdot 2\pi$~rad/s and $\omega_{1}=1.1\cdot 10^{9}\cdot 2\pi$~rad/s, which gives rise to $\Delta \omega=200\cdot 10^{6}\cdot 2\pi$~rad/s. Hence, the modulation frequency is about one fifth of the excitation frequencies. The static resistance $r_0=100$ Ohm and $\theta_1=\theta_{-1}=0$. 

Figures~\ref{fig1}(b)-(d) show the real and imaginary parts of the additional virtual impedance $Z^{\rm{v}}_{-1,1}=R^{\rm{v}}_{-1,1}+jX^{\rm{v}}_{-1,1}$ with respect to the modulation phase $\phi$, where $-1, 1$ refer to $\omega_{-1}$ and $ \omega_{1}$. From  Fig.~\ref{fig1}(b) we see that by tuning $\phi$, it is possible to realize pure resistive, pure reactive, or complex-valued virtual impedance. Furthermore, as $m$ or $a_{-1}/a_1$ increases, the real and imaginary parts of the virtual impedance grow. This is due to the fact that the virtual impedance is proportional to these parameters, as is obvious from Eqs.~\eqref{eq:intint} and \eqref{eq:intint1}. Thus, harnessing $\phi$, $m$, and $a_{-1}/a_1$ (or $a_1/a_{-1}$) indeed gives us an opportunity to fully control the virtual impedance. 

\begin{figure*}[t!]
	\centering
	\includegraphics[width=0.8\linewidth]{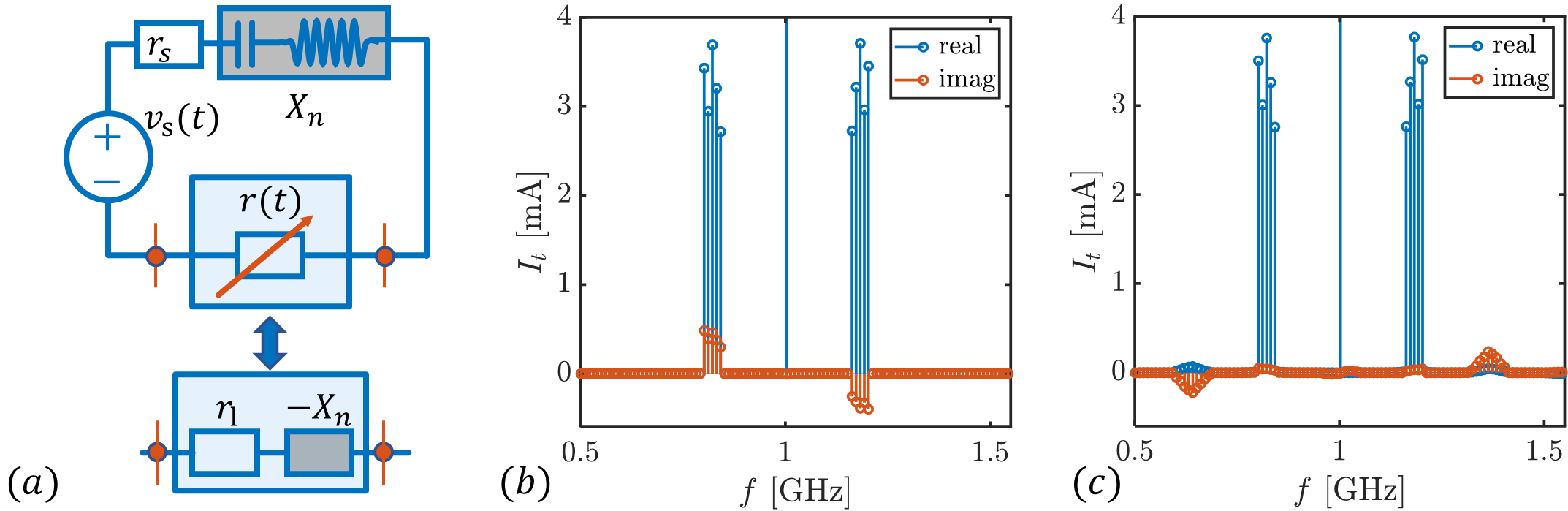}
	\caption{(a) Circuit schematic for voltage source $v_{\rm s}(t)$ connected to a resonant load with a time-varying resistance $r(t)$. (b) Current harmonics when modulation is off ($m_k=0$). (c) Current harmonics when modulation is on.}\label{Fig2}
\end{figure*}

Concerning the validity of the simplifying assumptions, let us consider two different cases shown in Figs.~\ref{fig1}(c) and (d). The first case corresponds to $m=0.25$ instead of $m=0.05$, while the second case corresponds to $a_{-1}/a_1=5$ instead of $a_{-1}/a_1=1$. We see that the analytical approximation is rather accurate for $m=0.05$, while for $m=0.25$ there is a slight disagreement which is more conspicuous in the real part. It is also seen that there is a good agreement between the analytical and numerical results even when $a_{-1}/a_{1}=5$. According to the above observation, we conclude that increasing $m$ plays a  more significant role in the accuracy of the analytical results in comparison with increasing the ratio $a_{-1}/a_{1}$.



Next, we leverage the virtual impedance introduced and explained to engineer the bandwidth of resonant circuits. The temporal modulation of a resistive element can be used to create a virtual impedance that compensates the reactive impedance of the resonant circuit at all input frequencies. As a result, the input power at all input harmonics is fully delivered to the load, as the load appears to be matched at all input frequencies. Let us consider the amplitude modulated input voltage signal
\begin{equation}
   v_\text{am}(t)=\cos{(\omega_{0}t)}\left[a_{c}+ \sum_{k=1}^{N} a_{s_k}\cos{(\Delta \omega_kt+\theta_{s_k})}\right],
   \label{Eq:V_AM 1} 
\end{equation}
where $\omega_{0}$ is the carrier angular frequency, $a_{c}$ is a real constant number, $N$ is the number of base-band signal harmonics, $a_{s_k}$ are the amplitudes, $\Delta \omega_k$ are the corresponding angular frequencies, and $\theta_{s_k}$ are the corresponding phases. We assume that the bandwidth of the base-band signal $\Delta \omega_N-\Delta \omega_1$ is smaller than $\Delta \omega_1-\omega_0$. 
It is useful to express $v_\text{am}$ as the summation of multiple frequency components as
 \begin{equation}
 v_\text{am}(t)=  \sum_{n=-N}^{N} a_{n}\cos{(\omega_{n}t+\theta_{n})},
\label{Eq:V_AM 2} 
\end{equation}
where $a_{\pm k}=0.5 a_{s_k}$, $a_0=a_c$, $-\theta_{-k}=\theta_{k}=\theta_{s_k}$, $\theta_{0}=0$, and $\omega_{\pm k}=\omega_{0}\pm \Delta \omega_k$. Let us assume that this voltage source is connected to the circuit shown in Fig.~\ref{Fig2}(a), where the total resistance is modulated in time according to the following rule:
\begin{equation}
   r_{\rm t}(t)=r_0+\sum_{k=1}^{N} r_0 m_{k} \cos{(\Delta \omega_k t+\phi_k)},
   \label{Eq:R(t) AM} 
\end{equation}
in which $r_0$ is the static resistive component that accounts for the static source resistance $r_{\rm{s}}$ and the static load resistance $r_{\rm{l}}$. Also, $m_{k}$ is the modulation amplitude, and $\phi_k$ is the modulation phase. Importantly, the modulation frequencies are the same as the base-band signal harmonics. The reactive elements $C$ and $L$ connected to the circuit resonate at $\omega_{0}$, and, as in any resonator, for frequencies $\omega_{k}$ the total reactive impedance is positive (inductive) $X_{k}$, and for frequencies $\omega_{-k}$ it is negative (capacitive) $X_{-k}$. For simplicity, we assume  that $|X_{-k}|=|X_{k}|$.

Choosing $\phi_k=\theta{s_k}+\frac{3\pi}{2}$ induces virtual impedances that can be expressed as
\begin{equation}
Z_{\pm k}^\text{v}=j X_{\pm k}^\text{v}=\mp j \frac{a_c}{a_{\pm k}} 0.5 m_{k} r_0 .
\label{Eq: Z_k^m} 
\end{equation}
The chosen modulation phase $\phi_k$ engineers the virtual impedances so that some virtual positive reactance is induced at frequencies $\omega_{+k}$, and some virtual negative reactance is induced at frequencies $\omega_{-k}$, or the sign of the induced reactance depends on their  position -- above or below -- with respect to the frequency $\omega_0$. By tuning $m_{k}$ and $a_c$ to satisfy $|X_{\pm k}^{\rm{v}}|=|X_{\pm k}|$, the circuit becomes purely resistive at all input frequencies, as the virtual reactances $X_{\pm k}^\text{v}$ compensate the reactances $X_{\pm k}$ stemming from the dispersive nature of the resonant circuit.


\begin{table}[b!]
  \begin{center}
    \caption{Parameters and impedance values}
    \label{tab:table1}
    \begin{tabular}{|c|c|c|c|c|c|c|c|c|}
    \hline
      $k$ & $\Delta f_k$ [MHz]  & $a_{s_k}$ & $X_{-k}$ & $X_{k}$ & $m_k$ & $X_{\pm k}^\text{v}$ & $Z_{-k}^\text{eq}$ & $Z_{k}^\text{eq}$\\
      \hline
      1 & 160 & 1.1 & -11 & 9.4 & 0.0224 & $\mp 10.18$ & 99.7-j0.9 & 99.5-j0.7 \\
      \hline
      2 & 170 & 1.3 & -11.7 & 9.9 & 0.0282 & $\mp 10.83$ & 99.7-j1.0 & 99.5-j0.8\\
      \hline
      3 & 180 & 1.5 & -12.5 & 10.5 & 0.0345 & $\mp 11.49$ &99.7-j1.1 & 99.5-j0.9\\
      \hline
      4 & 190 & 1.2 & -13.3 & 11 & 0.0292 & $\mp 12.15$ & 99.7-j1.2 & 99.5-j1.0\\
      \hline
      5 & 200 & 1.4 & -14.1 & 11.5 & 0.0359 & $\mp 12.82$ & 99.9-j1.3 & 99.5-j1.1\\
      \hline
    \end{tabular}
  \end{center}
\end{table}

As an example, we consider the voltage source $v_{\text{am}}$ having $\omega_0=1\cdot 10^9 \cdot 2\pi$~rad/s, $N=5$, and $\theta_{s_k}=0$. The values of $\Delta f_k=\frac{\Delta \omega_k}{2 \pi}$ and $a_{s_k}$ are shown in Table~1, where we assume that $a_{s_k}$ can take values between $1$ and $1.5$, and the values are chosen arbitrary. The other parameters are as following: $r_0=100$ Ohm, $\phi_k=\theta_{s_k}+\frac{3 \pi}{2}$, $C=5.07$ pF, $L=5$ nH, the resonance frequency is $\omega_0$, and the values of   $X_{\pm k}$ are shown in Table~1. To minimize $\sum_{k=1}^{N} m_k$, we choose $a_c=5$. In the above example, we assumed that $|X_{-k}|=|X_{k}|$, and in order  to perfectly match the circuit we needed to ensure that $|X_{\pm k}^{\rm{v}}|=|X_{\pm k}|$. This is not the case for this example because $|X_{-k}| \neq |X_{k}|$. Thus, to improve matching of the circuit we can choose $m_k$ that satisfies $|X^{\rm{v}}_{\pm k}|=\frac{|X_{-k}|+|X_{k}|}{2}$. This way we get
\begin{equation}
m_k=\frac{a_k}{a_{c}}\frac{|X_{-k}|+|X_{k}|}{ r_0}.
\label{Eq:m_k} 
\end{equation}

Values for $m_k$ and $X^{\rm{v}}_{\pm k}$ are calculated analytically and are  shown in Table~1 along with the numerically calculated total equivalent impedances $Z^{\text{eq}}_k={V_{\omega_{k}}\over I_{\omega_{k}}}$. 

\begin{figure*}
\centerline{\includegraphics[width=1\linewidth]{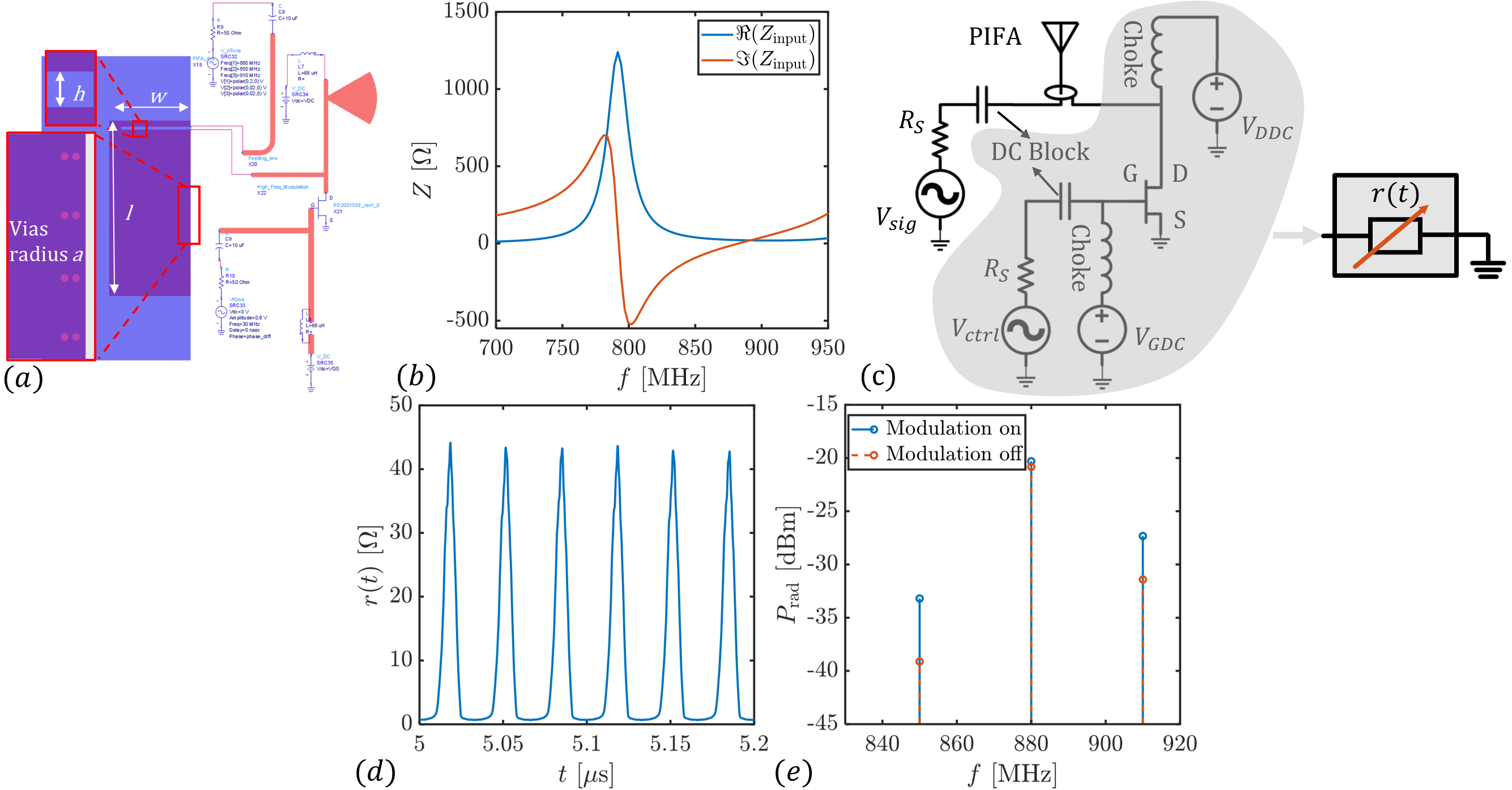}}
\caption{(a)--ADS model of the PIFA antenna where $h=1.5$~mm, $a=3$~mm, $w=39.32$~mm and $l=88.20$~mm. The PIFA is connected in series to the transistor. (b)--Simulation results for $Z$-parameters of the PIFA. (c)--Schematic of the whole system where the PIFA antenna is connected to a time-varying resistor. (d)--Profile of the time varying resistance. (e)--Comparison between radiated power $P_{\rm{rad}}$ by the PIFA antenna when modulation is on/off.}\label{Fig_PIFA}
\end{figure*}

As we see from Table~1, proper slow temporal modulation of the total resistance reduces the input reactance of the circuit. In this example we manage to keep the magnitude of all reactances smaller than $1.3$, which is a significant reduction compared to $X_{\pm k}$. The total current $I_{\rm t}(\omega)$ going through the circuit calculated numerically using the impedance-matrix method~\cite{Wang} is illustrated in Fig.~\ref{Fig2}. Figure~\ref{Fig2}(b)  shows the current harmonics when the circuit is static ($m_k=0$), and Fig.~\ref{Fig2}(c) shows the current harmonics in the time-modulated circuit. Due to modulation, the reactive components are reduced, and the active components are increased slightly. The modulation also results in some unwanted harmonics, but those that are inside our band of interest (from $\omega_{-5}$ to $\omega_{5}$) have significantly small amplitude, and the unwanted harmonics outside of this band are reactive and can be easily filtered out.




To verify the results experimentally, we study the introduced technique for a resonant load with a time-varying resistor. Specifically, the chosen test load is a typical planar inverted-F antenna (PIFA), and we connect it in series with a time-varying resistor. This way we can experimentally verify the validity of the proposed technique by measuring the power radiated by the PIFA antenna at all input frequencies of a modulated signal. For practical applications, temporal modulation of the source resistance can be used instead of adding an additional time-varying resistive component. Next, we present the antenna topology and the time-varying resistor realization, and show the numerical simulation results. 



The prototype antenna is a typical PIFA used in mobile devices. The structure of the antenna is shown in Fig.~3(a) (see Supplementary Material), and Fig.~3(b) shows the input impedance of the antenna between 700~MHz and 1000~MHz.
The realized time-modulated resistor (TMR) is based on a voltage-controlled transistor (MOSFET). 
The current-voltage (IV) transfer characteristics that determines the TMR performance is the most important factor to choose a proper transistor.  To realize a time-varying resistance, an AC control voltage is supplied over a DC gate bias  voltage. The control voltage amplitude is chosen so that the transistor operates in the linear regime. The proposed TMR topology is shown in Figs.~3(a) and (c). The time-varying resistance operation was first verified in simulations using Advanced Design System (ADS) software. Due to the high-frequency operation regime, the circuit is designed on a PCB with printed transmission lines to guide the signal and supply bias. In addition, a radial stub with a transmission-line section is used as a filter to prevent the high-frequency signal flow to the source supply. Figure~3(d) shows the gate-controlled resistance as a function of time, implemented by a PD20015SE MOSFET transistor. The values for the bias voltages are shown in Table~2. The TMR is nearly sinusoidal and varying with time at the frequency of 30 MHz, as required. The resistance of the transistor when modulation is off is equal to 2.45~Ohm.

\begin{table*}
    \centering
    \caption{Parameters and power level comparison}
    \label{tab:comparison}

\begin{tabular}{ |c|c|c|c|c|c|c|  }
\hline
    & Value &  & \multicolumn{2}{c|}{Power Level}\\
\hline
$V_{\rm DDC}$ & 1.3 V & Freq. & wo mod. & with mod. \\
\hline
$V_{\rm GDC}$ & 4.2 V & 850 MHz & -57.86 dBm & -55.33 dBm \\
\hline
$V_{\rm sig}$ & \begin{tabular}{c}
                  200 mV carrier freq. 880 MHz\\
                  20 mV AM modulation freq. 30 MHz\\
                 \end{tabular}  & 880 MHz & -31 dBm & -31 dBm \\
\hline
$V_{\rm ctrl}$ & 600 mV freq. 30 MHz &  910 MHz & -60.68 dBm & -57.10 dBm\\
\hline
\end{tabular}
\end{table*}

Figure~3(c) shows the connection of the PIFA antenna and the TMR circuit in series. The antenna is connected to the signal source $V_{\rm sig}$ in series with the TMR. The antenna terminals are connected between the drain terminal (D) of the MOSFET and the signal source, therefore, the antenna ground is not connected to the ground of the RF circuit. The source terminal (S) of the MOSFET is connected to the RF ground.
DC blocks and chokes are used to protect the source supplies. In this topology, the gate voltage consists of a DC voltage $V_{\rm GDC}$ to open the channel of the transistor and an AC voltage $V_{\rm ctrl}$ to control the varied resistance. For this prototype, the TMR is varied at frequency $30$~MHz and operates at a high frequency of $880$~MHz. The signal $V_{\rm sig}$ transferred to the antenna includes three harmonics: the carrier harmonic at 880 MHz and two side harmonics at 850 MHz and 910 MHz. The whole time-modulated load consisting of the antenna and TMR is simulated using ADS tools, as shown in Fig.~3(a). The values for voltages used in the simulation are shown in Table~2. 
Here, strong modulation is applied, as seen from Fig.~3(d). In addition, the voltage amplitudes ratio between the carrier and side harmonics is 10. As a result, the equivalent impedance of the transistor can be fully controlled by varying the modulation phase. For a specific modulation phase, the transistor's equivalent impedance at each side frequency is composed of some reactance and negative resistance. Hence, at this modulation phase the antenna matching is enhanced. Due to the many components added to the circuit, accurate calculation of the phase difference is challenging. Thus, the modulation phase is varied until the best performance is achieved.

We simulate the delivered power to the PIFA antenna in two cases: with and without time-varying resistance by turning the control voltage $V_{\rm ctrl}$ on and off. The power level comparison for these two cases is shown in Fig.~3(e). We observe  that the modulation significantly improves the power levels at the two side harmonics at 850/910 MHz. Furthermore, the carrier harmonic at 880 MHz is mostly kept unchanged, again in agreement with the theoretical expectations.
\begin{figure}[b!]
\centerline{\includegraphics[width=1\linewidth]{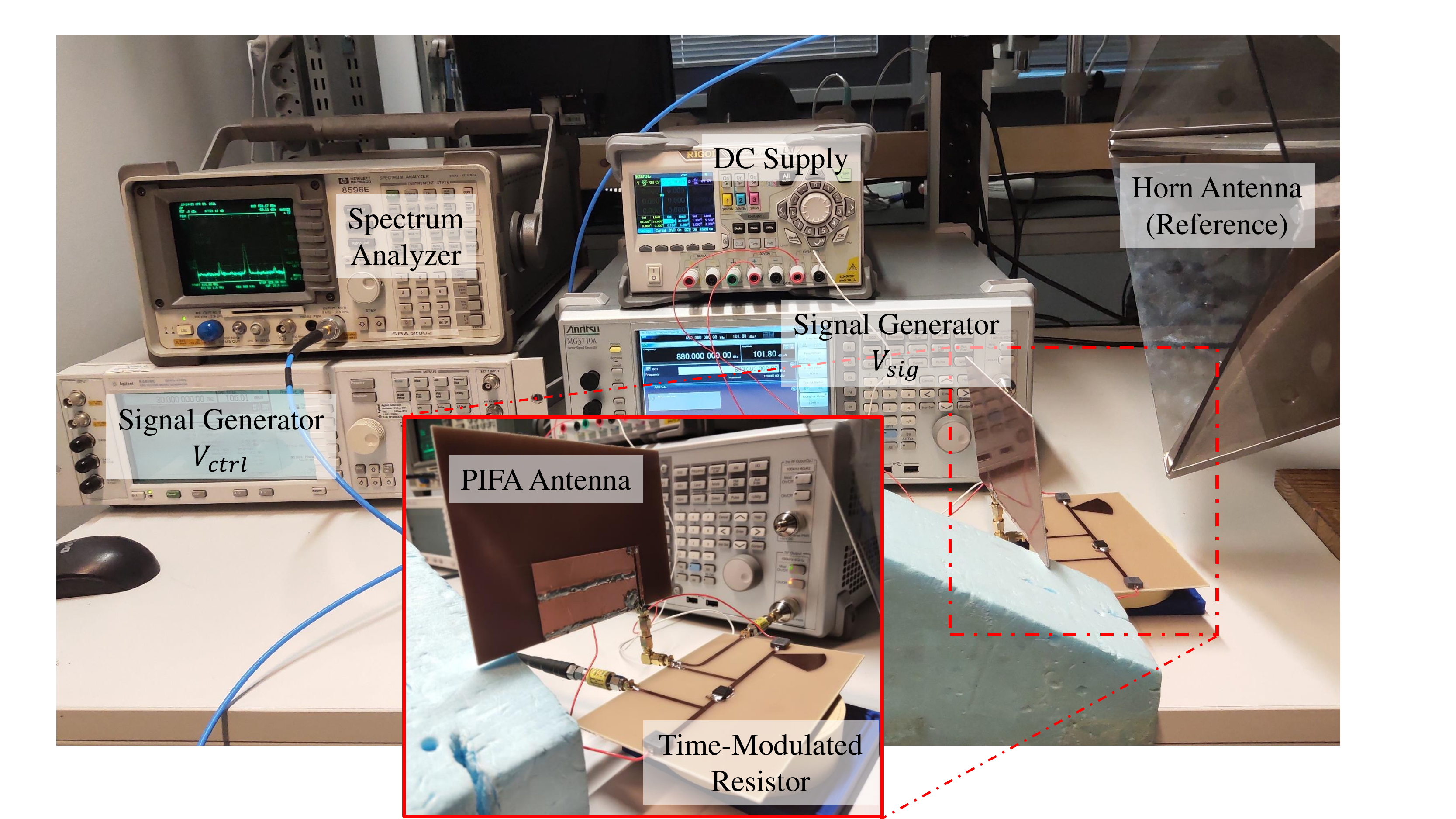}}
\caption{Measurement setup.}
\label{Fig_PIFA}
\end{figure}

\begin{figure}[b!]
\centerline{\includegraphics[width=1\linewidth]{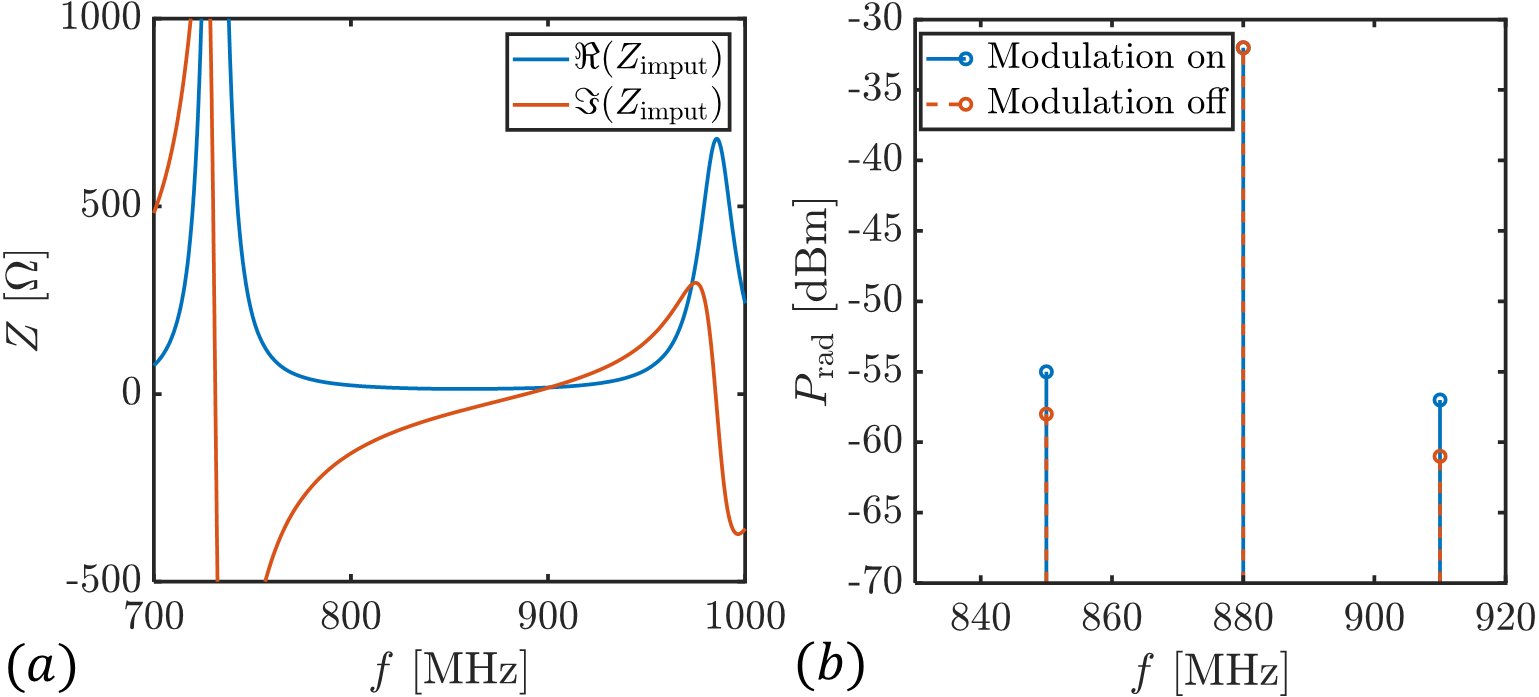}}
\caption{(a)--Measured $Z$-parameters of the PIFA. (b)--Comparison between measured radiated power $P_{\rm{rad}}$ by the PIFA antenna when modulation is on/off.}
\end{figure}

The fabricated antenna is shown in Fig.~4 (see Supplementary Material), while the measured input impedance is shown in Fig.~5(a). The measurement setup is shown in Fig.~4. Accurate measurements of the antenna current are  difficult at high frequencies. Thus, to validate the effect of the time-modulated resistor on the virtual matching of resonant antenna, a reference horn antenna connected to a spectrum analyzer was used as a receiving antenna to catch the power radiated  by the PIFA antenna. In addition, two signal generators are set up to generate the control voltage to the gate terminal and an amplitude modulated signal to the antenna. The generators are synchronized to allow control of the phase difference between the control voltage, or TMR's phase, and the signal phase. Finally, the transistor of the TMR circuit is biased by a DC voltage to the drain and gate terminals by a DC supply. All values used in the experiment are similar to the ones used in simulations, shown in Table~2.

Figure~5(b) shows the measured power levels at the carrier and two side harmonics that are received by the reference horn antenna. To measure these power levels of the harmonics  with and without modulation, the modulation signal generated by the controlled signal generator is alternately switched to the "ON/OFF" states. We find that for a certain value of the phase of the modulation signal, the side harmonics are enhanced by approximately $2-3$ dB, while the carrier harmonic is kept almost unchanged. The power levels are shown in Table~2. The experimental results confirm the concept of  using slow modulation of resistance for impedance engineering.

To realize the desired effect, we manually varied the modulation phase. Although the two generators are synchronized, there is no possibility to measure the actual value of the phase difference between the signal harmonics and the control voltage:  we can only vary this phase shift over the whole $360^\circ$  span. There are two different mechanisms for compensating the signal distortion due to antenna dispersion: creation of virtual reactances or virtual negative resistances at the side frequencies. As a result of our experiment we see the expected improvement, but we cannot clearly determine which effect dominates, the virtual reactance or virtual negative impedance. However, in any case, no power is pumped to the system. In the context of this work, a negative resistance is a model for the power coupled from other harmonics of the input signal~\cite{PhysRevApplied.17.064048}.


\label{sec:conclusions}

In this paper, we have presented a general approach to pass broadband modulated signals through narrowband resonant circuits (for example, small antennas) with minimized distortion and parasitic reflections.
The proposed solution is based on time modulations of resistance by an external bias that is controlled by the base-band signal. This can be realized either by adding a time-varying resistor in series with the source or by modulating the source resistance. 
The proposed solution is protected by a pending patent.


\begin{acknowledgments}
We wish to acknowledge the support of Huawei Finland and from the Academy of Finland under grant 330260. The authors wish to thank Dr. Mohammad S. Mirmoosa and Dr. Zlatoljub Milosavljevic for helpful discussions.
\end{acknowledgments}


\section*{Data Availability}

The data that support the findings of this study are available from the corresponding author upon reasonable request.


\section*{Conflict of Interest}

The authors have no conflicts to disclose.


\section*{References}

\nocite{*}
\bibliography{refrences}

\end{document}


\begin{center}
		{\bf
			{{Supplementary material for the paper}}\\
			{\Large{``Antenna bandwidth engineering through time-varying resistance''}}\\}
		{\normalsize{
				M.~H.~Mostafa, N.~Ha-Van, P.~Jayathurathnage, X.~Wang, G.~Ptitcyn, and Sergei~A.~Tretyakov}} 
		
	\end{center}
	\vspace{1cm}
	\tableofcontents
	\newpage

	
\section{PIFA antenna}

The antenna is printed on a grounded substrate. The left edge of the antenna is connected to the ground plane.  The substrate material is FR4 with the permittivity $\epsilon_{\rm r}=4.3(1-j0.025)$ and thickness $d=1.5$~mm. The resonant frequency is designed at 887~MHz. The PIFA antenna is modelled in ADS software. One can see that, in the studied frequency range, there are two resonances. The first resonance corresponds to a parallel RLC circuit. Around the second resonance, the input resistance is almost constant and the input reactance varies from capacitive to inductive, meaning that  the effective circuit is a resonant series RLC circuit.
The antenna radiates at this resonance and the radiation is mainly due to the equivalent magnetic current at the open edge.  According to the simulated input impedance, we determine the equivalent series lumped-parameter values ($R_{\rm r}$, $L$, and $C$) in the equivalent circuit. The equivalent lumped values  are $R_{\rm r}=24.48~\Omega$, $L=164 $~nH, and $C=0.195$~pF.

For the convenience of manually tuning the antenna, we made the antenna using copper tapes based on a metallized FR4 substrate. The effective circuit parameters of the fabricated antenna around the series resonance are fitted as $R_{\rm r}=15.48$ Ohm, $L=119 $~nH, and $C=0.27$~pF. One can see that the measured antenna parameters are very close to the simulated results, exhibiting both parallel and series resonances in the studied frequency range. However, the distance between the parallel and series resonances of the tested antenna is larger than for the simulated one. This is because in the test antenna, we tune the slot length and deliberately move the parallel resonance further away compared to the simulated model. By doing this, the antenna can be more precisely modelled by a series circuit around $880 $~MHz since the coupling between series and parallel resonances is reduced.
The effective circuit parameters of the test antenna around the series resonance are fitted as $R_{\rm r}=15.48~\Omega$, $L=119 $~nH, and $C=0.27$~pF.
	

\section{Reference measurement}

As a reference, we have measured the spectrum of the signal supplied by the generator, directly connecting its output cable to the spectrum analyser. We observe the differences between the power levels of side harmonics to the carrier in the initial signal supplied by the power generator. Then, we compare these differential power levels to the differential power levels of the signal received by the horn antenna when the modulation device is on or off. The differential power levels are shown in Table~S1.

We note that the input reference signal is not ideal: the side harmonics have somewhat different amplitudes although it is supposed to be a single-harmonic amplitude modulated signal. This brings an additional uncertainty in measurements. We conclude that within the measurement uncertainties the device works as expected. Differences in the effects at the lower and higher frequencies are explained by an asymmetric frequency dispersion of the antenna: the two side bands are distorted differently, so the effect of the time-modulated resistor is also somewhat asymmetric.

\begin{table*}
    \centering
    \caption{Differential power levels}
    \label{tab:comparison}
\begin{tabular}{ |c|c|c|c|  }
\hline
frequency & reference & wo mod. & with mod. \\
\hline
850 MHz & 26.7 dB & 26 dB & 24 dB\\
\hline
910 MHz& 27.3 dB & 29 dB & 26 dB\\
\hline
\end{tabular}
\end{table*}

